\documentclass[12pt,preprint]{aastex}

\slugcomment{Astrophysical Journal, in press}
\shorttitle{Photon Bubbles Around Young Massive Stars}
\shortauthors{Turner et al.}

\newcommand{\divv}{{\bf\nabla\cdot v}}

\begin{document}
\title{Photon Bubbles in the Circumstellar Envelopes\\ of Young
Massive Stars}

\author{N.~J.~Turner\altaffilmark{1},
E.~Quataert\altaffilmark{2} and
H.~W.~Yorke\altaffilmark{1}}

\altaffiltext{1}{Jet Propulsion Laboratory, MS 169-506, California
Institute of Technology, Pasadena CA 91109; neal.turner@jpl.nasa.gov}
\altaffiltext{2}{Astronomy Department, University of California at
Berkeley, 501 Campbell Hall, Berkeley, CA 94720}

\begin{abstract}
  We show that the optically-thick dusty envelopes surrounding young
  high-mass stars are subject to the photon bubble instability.  The
  infrared radiation passing through the envelope amplifies
  magnetosonic disturbances, with growth rates in our local numerical
  radiation MHD calculations that are consistent with a linear
  analysis.  Modes with wavelengths comparable to the gas pressure
  scale height grow by more than two orders of magnitude in a thousand
  years, reaching non-linear amplitudes within the envelope lifetime.
  If the magnetic pressure in the envelope exceeds the gas pressure,
  the instability develops into trains of propagating shocks.
  Radiation escapes readily through the low-density material between
  the shocks, enabling accretion to continue despite the Eddington
  limit imposed by the dust opacity.  The supersonic motions arising
  from the photon bubble instability can help explain the large
  velocity dispersions of hot molecular cores, while conditions in the
  shocked gas are suitable for maser emission.  We conclude that the
  photon bubble instability may play a key role in the formation of
  massive stars.
\end{abstract}


\section{INTRODUCTION\label{sec:intro}}

Massive stars are important for the formation and destruction of
molecular clouds and the chemical enrichment of galaxies, but our
understanding of their origins remains limited.  The high luminosities
of young massive stars mean that radiation forces can hinder the
accretion of surrounding material required for growth.  Although stars
of over 100~M$_\odot$ are known (e.g. Figer~2005),
the Eddington limit for radiation pressure on dust is exceeded at
about 30~M$_\odot$ in the simplest case of spherical symmetry with a
Solar abundance of interstellar grains \citep{wc87}.  Higher masses
are reached if the symmetry is broken by the presence of an accretion
disk \citep{na89,ja96,ys02} or a large-scale radiation-inflated cavity
\citep{kk05}.  Further evidence against a spherical geometry comes
from the result that clumpiness in the obscuring material around young
massive stars can explain the observed combination of high
near-infrared brightness with large line-of-sight extinctions, as well
as the variety in 10-$\mu$m silicate features, while spherically
symmetric obscuration cannot \citep{iw06}.  An alternative way to
overcome the Eddington limit is by the ram pressure of the accreting
material at sufficiently high mass flow rates.  The high accretion
rates can arise in parent molecular cloud cores with turbulent
velocity dispersions greater than 1~km~s$^{-1}$ \citep{wc87,mt03} and
this condition is met in many observed hot cores \citep{se03}.  The
mechanism driving the movements in the molecular gas is unclear.

Maser emission associated with many massive young stars is a useful
probe of the kinematics and physical conditions in the surrounding
gas.  VLBI mapping of water masers shows the emission typically comes
from multiple spots each about 1~AU in size \citep{gw94}.  Modeling
indicates the population inversion can occur in both 100~km~s$^{-1}$
dissociative J-shocks \citep{eh89} and 10~km~s$^{-1}$ non-dissociative
C-shocks \citep{kn96}.  The inferred amplification paths are longer
than the spot sizes, suggesting the fronts are extended along the line
of sight.  The strongest and least variable emission features
typically are at velocities within 8~km~s$^{-1}$ of the parent cloud
\citep{bc05}.  Zeeman measurements indicate the masing gas has
line-of-sight magnetic fields of 0.01--0.6~Gauss
\citep{fg89,st01,st02,vd06} with pressures similar to the turbulent
pressure and in some cases exceeding the gas pressure.

Here we discuss a new candidate process for separating the outgoing
radiation from the inflowing mass in regions of massive star
formation.  Photon bubbles are a local linear overstability of the
magnetized, radiation-supported atmospheres in neutron star accretion
columns \citep{ar92} and black hole accretion disks \citep{ga98}.
Photon bubbles are also found at the photosphere in main-sequence
O-stars \citep{ty04} if the magnetic fields are as strong as that
measured in $\theta^1$~Orionis~C by \cite{db02}.  The instability
draws energy from the background photon flux as follows.  A
compressive MHD wave propagating at an angle to the flux causes gas
density variations.  Greater radiation fluxes pass through the
low-density regions, causing forces that accelerate the gas along the
magnetic field lines.  The low-density regions are further evacuated,
increasing the amplitude of the wave.  The stronger density variations
in turn mean larger flux perturbations, leading to runaway growth.  A
diagram showing the mechanism is in Turner et al. (2005; their
figure~7).

Favorable conditions for photon bubble instability are found in the
dusty envelopes immediately surrounding the ultracompact H{\sc II}
regions of young high-mass stars.  The optical depth is greater than
unity, allowing perturbations in the radiative flux, but also small
enough that photons diffuse through in much less than the sound
crossing time, so that the radiation is not trapped inside the
disturbances.  Large density variations are possible because the
photon and magnetic pressures are similar to or greater than the gas
pressure.  Due to the dust absorption opacity, gas and radiation reach
mutual thermal equilibrium much faster than the periods of the
candidate waves.  Under these conditions the fastest-growing photon
bubble modes have wavelengths shorter than the gas pressure scale
height $c_i^2/g$.  The asymptotic growth rate $g/(2c_i)$ depends on
the isothermal gas sound speed $c_i$ and the gravitational
acceleration $g$ \citep{bs03}.  On reaching non-linear amplitudes, the
photon bubbles develop into trains of propagating shocks \citep{be01}.

We have made numerical radiation-MHD calculations of a small portion
of the envelope around a young massive star to explore the effects of
photon bubbles.  The equations and numerical methods are described in
\S\ref{sec:eqns}, the domain and initial conditions in \S\ref{sec:ic}.
The linear growth of photon bubbles is demonstrated in
\S\ref{sec:linear} and the development of shocks in
\S\ref{sec:shocks}.  The implications of the results are discussed in
\S\ref{sec:disc} and our conclusions are summarized in
\S\ref{sec:conc}.

\section{EQUATIONS AND METHOD OF SOLUTION\label{sec:eqns}}

The standard equations of radiation MHD are solved using the {\sc
Zeus} code \citep{sn92a,sn92b} with its flux-limited photon diffusion
module \citep{by90,ts01}.  The radiation field is averaged in
frequency and angle, leading to
\begin{equation}\label{eqn:cty}
{D\rho\over D t}+\rho\divv=0,
\end{equation}
\begin{equation}\label{eqn:gasmomentum}
\rho{D{\bf v}\over D t} = -{\bf\nabla}p
	+ {1\over 4\pi}({\bf\nabla\times B}){\bf\times B}
        + {1\over c}\chi\rho{\bf F} - \rho{\bf g},
\end{equation}
\begin{equation}\label{eqn:radenergy}
\rho{D\over D t}\left({E\over\rho}\right) =
	- {\bf\nabla\cdot F} - {\bf\nabla v}:\mathrm{P}
	+ \kappa\rho(4\pi B - c E),
\end{equation}
\begin{equation}\label{eqn:gasenergy}
\rho{D\over D t}\left({e\over\rho}\right) =
	- p\divv - \kappa\rho(4\pi B - c E),
\end{equation}
\begin{equation}\label{eqn:radmomentum}
{\bf F} = -{c\Lambda\over\chi\rho}{\bf\nabla}E,
\end{equation}
and
\begin{equation}\label{eqn:induction}
{\partial{\bf B}\over\partial t} = {\bf\nabla\times}({\bf v\times B})
\end{equation}
\citep{mm84}.  The quantities $\rho$, ${\bf v}$, $e$, and $p$ are the
gas density, velocity, internal energy density, and pressure,
respectively, while ${\bf B}$ is the magnetic field and ${\bf g}$ the
acceleration due to the stellar gravity.  The gas and radiation have
distinct temperatures.  The photons are represented by their
bolometric energy density $E$, energy flux ${\bf F}$, and pressure
tensor $\rm P$.  The absorption opacity $\kappa$ and total opacity
$\chi$ are due to dust well-mixed with the gas and are both taken
equal to 2~cm$^2$~g$^{-1}$ in the relevant temperature range near
$1\,000$~K \citep{sh03}.  The flow cools by emitting photons from the
dust at a rate proportional to the blackbody value $B=\sigma_B
T_g^4/\pi$, where $\sigma_B$ is the Boltzmann constant,
$T_g=p\mu/({\cal R}\rho)$ the temperature of the gas and grains,
$\mu=2.34$ atomic mass units the mean mass per particle and ${\cal R}$
the gas constant.  In equation~\ref{eqn:radmomentum}, the flux-limiter
$\Lambda$ is equal to $1/3$ in optically-thick regions.  Causality is
preserved in regions where radiation energy density varies over
optical depths less than unity, by reducing the limiter toward zero
using the prescription of Levermore \& Pomraning (1981; their
equation~22).
An implicit differencing scheme is used for the radiation terms,
allowing numerical stability with timesteps longer than the diffusion
step $(\Delta x)^2\chi\rho/(2c\Lambda)$ which is proportional to the
square of the grid spacing $\Delta x$.  The equations are closed by
assuming an ideal gas equation of state $p=(\gamma-1)e$, with
$\gamma=7/5$ for molecular hydrogen.

\section{DOMAIN AND INITIAL STATE\label{sec:ic}}

The starting conditions chosen are appropriate to the dusty envelope
surrounding the ultracompact H{\sc II} region of a young O6 star and
are based on results from a frequency-dependent radiation hydrodynamic
calculation of the collapse of a 60~M$_\odot$ molecular clump
\citep{ys02}.  The domain for our calculations is a small patch near
the magnetic equator or pole of the clump.  The patch lies outside the
dust sublimation radius and the stellar radiation is assumed to have
been reprocessed to infrared wavelengths by the intervening material.
As we wish to explore the consequences of photon bubbles for the
Eddington limit, force balance is assumed between gravity, radiation
and gas pressure forces.  The patch is centered 400~AU from the star
and extends 40~AU either side of the center along the two Cartesian
coordinates $(x, z)$.  A uniform acceleration equal to the gravity $g$
of the 33.6~M$_\odot$ star at domain center is applied along the
negative $z$-direction.  The stellar luminosity of $2\times
10^5$~L$_\odot$ leads to an upward radiation force that is 92\% of the
gravitational force.  The density and temperature at domain center are
$10^{-14}$~g~cm$^{-3}$ and $1\,000$~K, respectively, and the remainder
of the domain is placed in radiative flux balance and thermal and
hydrostatic equilibrium.  The boundaries are treated as follows.  The
lower edge is impermeable to the gas and the radiation temperature
there is fixed at its initial value.  A periodic condition is applied
across the side boundaries, and the top edge allows outflow but no
inflow.  The magnetic field is uniform with a strength 0.1~Gauss and
the radiation, magnetic and gas pressures have the ratio $7.1 / 1.1 /
1$ at domain center.  The radiation and isothermal gas acoustic speeds
are $c_r=(4E/9\rho)^{1/2}=5.8$~km~s$^{-1}$ and
$c_i=(p/\rho)^{1/2}=1.9$~km~s$^{-1}$ and the Alfv\'{e}n speed is
$v_A=|{\bf B}|/\sqrt{4\pi\rho}=2.8$~km~s$^{-1}$.  The corresponding
radiation and gas pressure scale heights are $c_r^2/g = 180$~AU and
$c_i^2/g = 19$~AU.  Internal gradients are slight because the domain
is smaller than the radiation pressure scale height: the temperatures
and densities at the top and bottom boundaries differ from those at
the center by at most 12\%.  The domain is optically-thick and the
initial optical depth through the height and width is 24.

Under these conditions, vigorous photon bubble instability is expected
since the radiative flux exceeds the threshold $c_iE$ \citep{bs03} by
more than a thousand times.  The fastest-growing modes are slow
magnetosonic waves with fronts separated by less than the gas pressure
scale height.  In the short-wavelength limit, the linear analysis
indicates a minimum $e$-folding time over all magnetic field
orientations of $2c_i/g = 100$~yr.  However, extremely
short-wavelength perturbations with optical depths less than unity
hardly disturb the radiative flux and will grow slowly or not at all,
so the fastest-growing modes are expected to fit no more than
24~wavelengths between the boundaries.  Gas, grains and radiation will
remain at a common temperature in the waves because the oscillations
are much slower than the emission and absorption of radiation.  The
rate-limiting process in the thermal equilibration is the exchange of
heat in collisions between gas molecules and grains.  The time for
these two components to reach a shared temperature \citep{bh83} is
about $0.1$~yr.  Photons also diffuse quickly, spreading across the
domain in $0.1$~yr so that acoustic disturbances in the wavelength
range of interest are not supported by radiation pressure and travel
at the gas sound speed $c_i$.  The gas and grains are dynamically
well-coupled by drag forces as the stopping time for sub-micron grains
is less than $0.01$~yr.

The evolution of the magnetic fields is described by ideal MHD since
there is sufficient ionization for good coupling to the gas.  The
ionization includes contributions from both cosmic rays and collisions
of neutrals with readily-ionized metals such as potassium.  We
estimate the cosmic ray ionization by solving a chemical reaction
network including dissociative recombination of molecular hydrogen,
charge exchange with metal atoms and radiative recombination of metal
ions \citep{od74,in06a}, using the interstellar cosmic ray ionization
rate of $10^{-17}$~s$^{-1}$ and choosing magnesium with Solar
abundance as a representative metal.  Recombination on grain surfaces
can be neglected because the ions and electrons remain in the gas
phase at the high temperatures found here \citep{um83}.  The resulting
equilibrium electron fraction is $6\times 10^{-8}$.  After an abrupt
doubling of the density as on passing through a shock, the ionization
fraction falls slightly to $4\times 10^{-8}$ on a time scale of
$70$~yr.  Solving the Saha equation for potassium with Solar abundance
indicates thermal ionization produces a smaller contribution to the
electron fraction of $3\times 10^{-10}$.  Under these conditions, the
strongest non-ideal effect is ambipolar drift.  The ratio of the
ambipolar drift term to the induction term in the magnetic field
evolution equation depends on the flow speed and the characteristic
scale of field gradients \citep{ss02a}.  The typical speed in the
photon bubbles is the gas sound speed and ambipolar diffusion is
important only for magnetic fields changing over scales shorter than
0.03~AU.  The induction term dominates for the fast-growing photon
bubble modes since these have optical depth per wavelength greater
than unity, corresponding to scales longer than 3~AU.

To check that the initial hydrostatic equilibrium is long-lived we
make a series of radiation hydrodynamic test calculations without
magnetic fields.  An upper bound on the numerical error is obtained by
dividing the domain into a grid of just $32\times 32$~zones.  At the
start of each calculation, a random 1\% density perturbation is
applied in every zone in the middle half of the domain height.  In the
first case, the timestep is restricted to no longer than five times
the diffusion step.  Under the implicit differencing scheme used for
the radiation terms, the integration is numerically stable with the
longer timestep but the accuracy is reduced.  Vertical oscillations
occur because the initial state is an equilibrium solution of the
continuum hydrostatic equations and differs slightly from the
equilibrium solution of the finite difference equations.  Mass is lost
through the top boundary during the upward excursions, leaving the
remainder further from hydrostatic balance.  The vertical oscillations
grow over time and half the mass is lost in 3900~yr.  If the top
boundary is made impermeable, so that no mass is lost, the
oscillations instead damp away.  However because we wish to allow
material to leave the domain, we instead reduce the timestep to twice
the diffusion step, leading to higher numerical accuracy.  The
oscillations then start with a smaller amplitude and damp over time
rather than growing, and the RMS vertical speed remains less than
0.2\% of the gas sound speed through the end of the calculation at
$10^4$~yr.  Results are similar in a further calculation with
timesteps equal to the diffusion step.  All the calculations described
in the remainder of the paper use timesteps no longer than twice the
diffusion step.  The computational load is heavy because at the
typical spatial resolution of $64\times 64$~zones there are initially
$10^8$ diffusion steps per thousand years.  The execution time is
proportional to the inverse fourth power of the grid spacing, since
both the number of zones and the number of timesteps scale as $(\Delta
x)^{-2}$.

\section{LINEAR GROWTH OF PHOTON BUBBLES\label{sec:linear}}

We first compare the growth rates of small-amplitude photon bubbles
against the results of a local linear WKB analysis by \cite{bs03}.
Solutions of the dispersion relation (their equation~49) indicate the
group velocity for the fastest modes is approximately parallel to the
magnetic fields.  To reduce any effects of the waves interacting with
the lower boundary, we start with the magnetic field horizontal and
apply random density perturbations to every grid zone in the top half
of the domain height.  For accurate measurements over a long growth
phase, small perturbations are chosen with an amplitude of just one
part per million.

\begin{figure}[tb!]
\epsscale{0.8}
\plotone{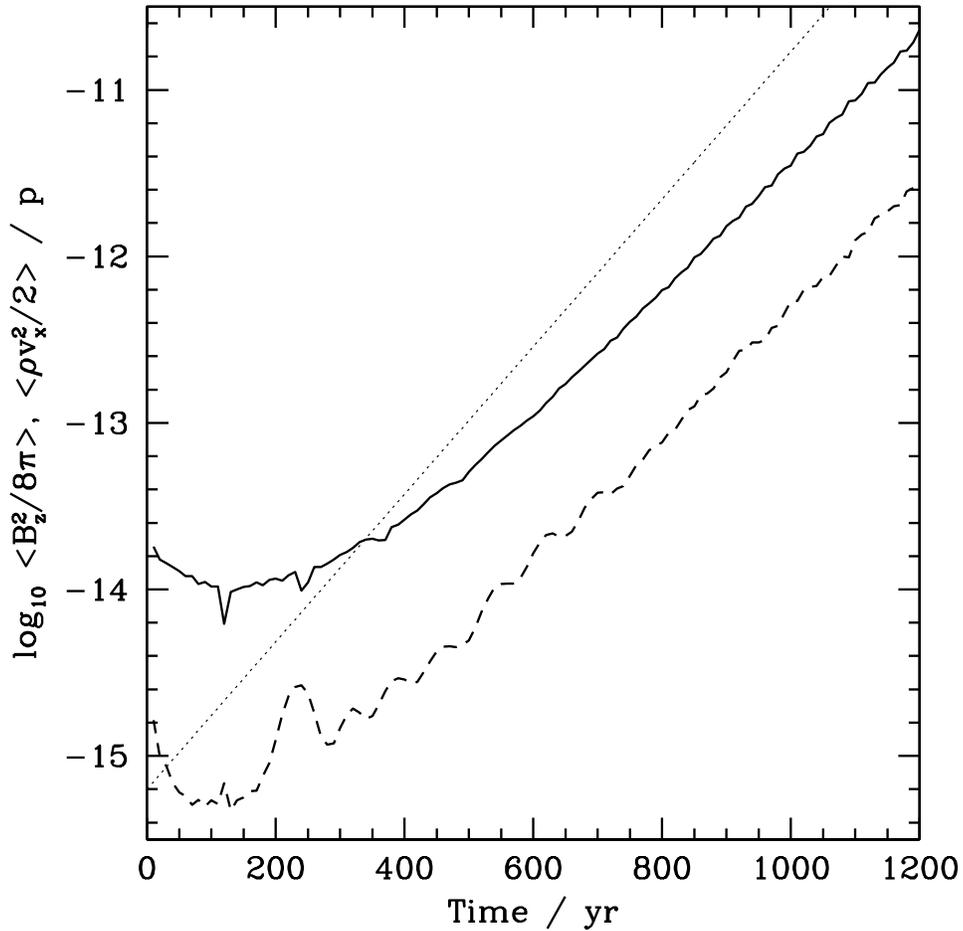}
\caption{\small Mean energy densities versus time in the
  numerical calculation with $64^2$~zones.  The solid line shows the
  kinetic energy in horizontal motions and the dashed line the energy
  in the vertical component of the magnetic field.  Both are divided
  by the initial gas pressure at domain center.  The growth rate of
  the fastest mode in the linear analysis is indicated by a dotted
  line.
  \label{fig:growthtime}}
\end{figure}

\begin{figure}[tb!]
\epsscale{0.4}
\plotone{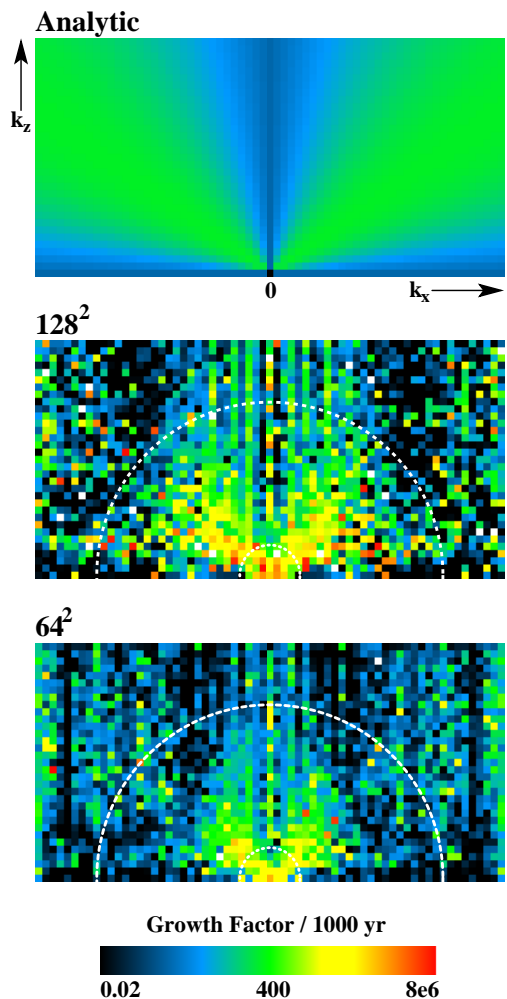}

\caption{\small
  Growth rate of individual Fourier components versus horizontal and
  vertical wavenumber.  Solutions of the linear dispersion relation
  for domain center are at top.  Results from numerical radiation-MHD
  calculations with $128^2$ and $64^2$~zones are in the middle and
  bottom panels.  Zero wavenumber lies at bottom center in each panel
  and the modes furthest from the origin have 32~wavelengths in the
  domain width and height.  Colors show the amplification per
  $1\,000$~yr on a logarithmic scale.  Small dotted arcs mark the
  inverse gas pressure scale height and larger dashed arcs the
  wavenumbers having unit optical depth.
  \label{fig:growthrate}}
\end{figure}

After an initial transient lasting a few hundred years, the
disturbance is dominated by exponentially growing modes.  The energy
in the perturbations increases over time as shown in
figure~\ref{fig:growthtime}.  The measured growth rate of 4.5 velocity
$e$-foldings per $1\,000$~years is slightly less than the 5.1
$e$-foldings of the fastest mode in the linear analysis.  Growth rates
are also measured using the spatial Fourier power spectra of the
horizontal component of the velocity.  The amplitudes after several
$e$-foldings are divided by those at the onset of linear growth to
obtain plots of the growth rate versus horizontal and vertical
wavenumber as in figure~\ref{fig:growthrate}.  Growth is quickest for
modes propagating about $40^\circ$ from the horizontal, as expected
based on the linear analysis.  An adequate spatial resolution of ten
or more grid zones per wavelength is reached for modes within
12~pixels of the origin in the calculation with $128^2$~zones, and for
modes within 6~pixels of the origin in the case with $64^2$~zones.
The linear analysis indicates the fastest-growing modes have
wavenumbers greater than the inverse gas pressure scale height.  Both
of the radiation-MHD calculations adequately resolve some modes lying
in the range of rapid growth, while the higher wavenumbers are
numerically damped.  The results in figure~\ref{fig:growthrate}
include vertical strips of high growth rate at small horizontal
wavenumbers.  These correspond to global oscillations of the
stratified background state that vary in amplitude on the photon
diffusion timescale with no net growth over longer periods.

\clearpage
\clearpage

\clearpage

\section{DEVELOPMENT OF SHOCKS\label{sec:shocks}}

\begin{figure}[tb]
\epsscale{0.6}
\plotone{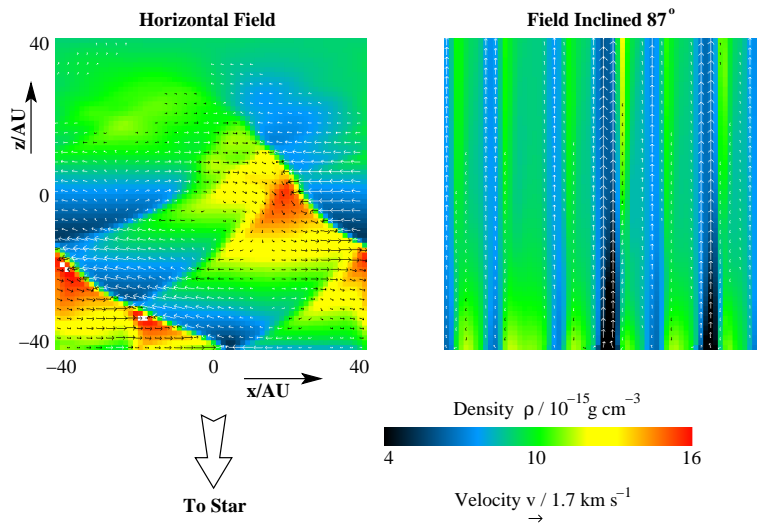}
\caption{\small Trains of shocks resulting from the photon bubble
  instability acting on initial random 1\% density perturbations.  At
  left the starting magnetic field is horizontal and the time shown is
  $2\,100$~yr.  At right the field is initially $87^\circ$ from
  horizontal and the time shown is $1\,300$~yr.  Colors indicate the
  density on a linear scale and arrows the velocity.  Animations showing
  the development of the shock trains are available in the online
  edition of the journal.
  \label{fig:shocktrains}}
\end{figure}

\begin{figure}[tb]
\epsscale{0.6}
\plotone{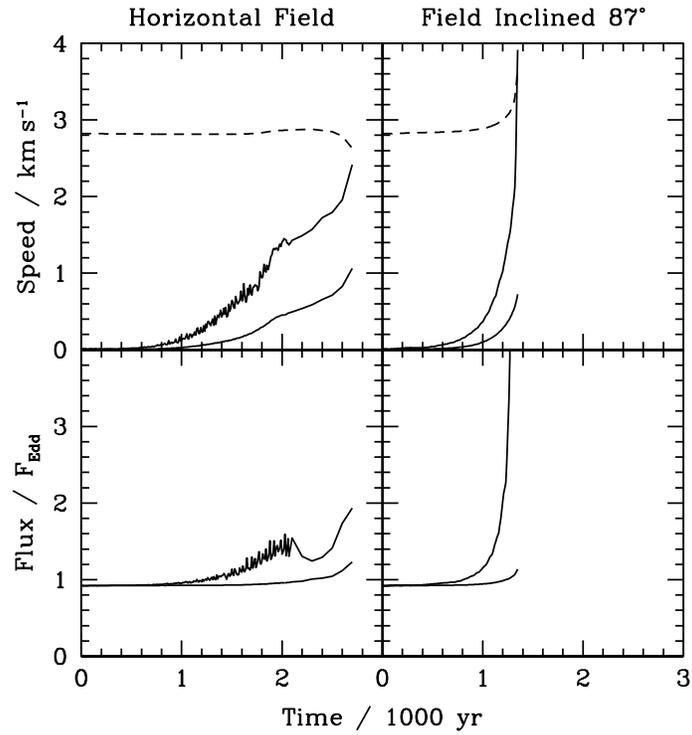}
\caption{\small Gas flow speed (top panels) and the vertical component
  of the radiation flux (bottom panels) versus time, in the
  calculations with magnetic fields initially horizontal (left) and
  inclined $87^\circ$ (right).  The upper solid curve in each panel
  shows the maximum value and the lower curve shows the RMS speed or
  mean flux.  Dashed lines in the top panels indicate the RMS Alfven
  speed.
  \label{fig:veldiseddlim}}
\end{figure}

The molecular cores where young stars form are certainly inhomogeneous
at levels greater than 1\%.  Such disturbances are expected to grow to
order unity in less than 800~yr for the fastest modes found in
section~\ref{sec:linear}.  We apply initial random 1\% density
perturbations to every grid zone in the middle half of the domain
height, in calculations with $64^2$~zones and magnetic fields (1)
horizontal and (2) inclined $87^\circ$.  The exponential growth of the
photon bubbles is faster in the case with the inclined fields because
the radiative flux perturbations are parallel to the gas velocities
and have the maximum driving effect \citep{tb05}.  On reaching large
amplitudes the propagating waves steepen into shocks
(figure~\ref{fig:shocktrains}).  The orientations of the shocks at
first match those of the linear wave fronts.  The gas oscillates back
and forth along the magnetic field lines, driven by the perturbations
in the radiation force.  Because the gas moves parallel to the fields,
the field strength and orientation are basically unchanged by the
shocks \citep{be01}.  The temperatures in the flow differ by less than
1\% from the starting values, as the emission and diffusion of photons
are much faster than the gas travel time from one shock to the next.
The shocks generally propagate at different speeds, merging when they
coincide so that the strength and separation of the remaining fronts
increase and the radiative flux passing through the flow also rises.
Eventually a single shock dominates in both calculations.  Further
growth is limited by the small size of the domain.  The
horizontal-field calculation terminates at $2\,700$~yr when the single
remaining shock is strong enough to bend the field lines across the
top boundary, leading to the escape of much of the mass.  The
$87^\circ$-field calculation ends at $1\,350$~yr when the flow rising
from the bottom boundary, upstream of the strongest shock, leaves
behind a density orders of magnitude lower than the surroundings and
the diffusion timestep becomes infeasibly short.  The velocity
dispersion and radiation flux generally increase with time
(figure~\ref{fig:veldiseddlim}), reaching about 1~km~s$^{-1}$ and
exceeding the Eddington limit, respectively, around the time the
calculations end.  The velocity difference across the shocks at this
time is comparable to the Alfven speed of about 3~km~s$^{-1}$.  The
excess radiation force is sufficient to accelerate all the material
out the domain top in the horizontal-field case within $2\,000$~years,
if the smooth initial mass distribution were still in place.  The fact
that 97\% of the material remains on the grid at this time is due to
the easy diffusion of the photons through the low-density gas between
the shocks.

\clearpage

\clearpage

As a check on the numerical accuracy during the non-linear evolution,
we made two versions of the horizontal-field calculation using
timesteps 1 and 2 times the diffusion step, with a lower spatial
resolution of $32^2$~zones.  Similar results are obtained in both.  At
$2\,300$~yr, the shock fronts are offset by about two grid cells and
elsewhere the densities differ less than 10\%.  These results indicate
that a timestep twice the diffusion step is sufficient to capture the
main features of the shock trains.

The growth of shocks in accreting material is examined using an extra
calculation with a shallower gas pressure gradient, such that the
initial radiation, gas and gravitational forces yield a net
acceleration toward the star 1\% that of gravity.  The calculation is
carried out in the reference frame of the infalling gas, using
$32^2$~zones and parameters otherwise matching the horizontal-field
run shown in figure~\ref{fig:shocktrains}.  The results are almost
identical to the version initially in force balance: the growth of the
shock trains allows the radiative flux to increase with little change
in the mean vertical radiation force.  The maximum flux exceeds the
Eddington limit at $1\,184$~yr while in the corresponding calculation
initially in force balance the limit is reached at $1\,210$~yr.  Both
results are very similar to figure~\ref{fig:veldiseddlim}, lower left
panel.  Despite the super-Eddington fluxes, the gas continues falling
toward the star until the run again ends when the motions grow strong
enough to bend the field lines.

\section{DISCUSSION\label{sec:disc}}

The analytic and numerical results show photon bubbles can grow
quickly in the envelope around an O6 star.  We next estimate the
variation with stellar luminosity.  The criterion for instability,
$F>c_iE$, is exceeded by a factor of a thousand just outside the grain
sublimation distance in the cases considered above.  The three
quantities involved vary little with luminosity: the gas sound speed
$\sqrt{{\cal R}T_s/\mu}$ and radiation energy density
$4\sigma_BT_s^4/c$ are fixed by the sublimation temperature $T_s$,
while the flux at the sublimation point corresponds to a balance
between the grains' absorption and re-emission of stellar radiation
\citep{wc86} and is nearly independent of luminosity if the dust-free
interior gas is optically thin.  The instability criterion applies
provided the density is high enough so the material is optically thick
to its own radiation, but not so high that the gas pressure exceeds
the magnetic or radiation pressure.  If these conditions are met, the
instability can occur in the circumstellar envelopes of stars with a
wide range of luminosities.

The growth time also is roughly luminosity-independent at the
sublimation point and increases with distance from the star in almost
the same way as the free-fall time.  The minimum time per $e$-folding
is approximately $2c_i/g_{\rm eff}$ \citep{bs03}, whether the gas is
stationary or infalling.  The effective gravity $g_{\rm eff}$ in the
gas frame is the radiative acceleration $\chi F/c$ if radiation is the
largest outward force and gas pressure gradients are unimportant.  The
minimum growth time at the sublimation distance is then approximately
$2cc_i/(\chi F)$ and depends little on the luminosity.  Outside the
sublimation point, the growth time increases with the distance from
the star $r$ approximately as $r^{1.6}$, since the gas sound speed
$c_i$ varies as $r^{-0.4}$ based on the temperature profile in the
optically-thick part of an envelope in spherical symmetry
\citep{wc86}.  The free-fall time scales similarly, being proportional
to $r^{3/2}$.  We conclude that photon bubbles are relevant throughout
the region that is optically-thick and coupled to the magnetic fields.
Shocks can develop if the time for gas to flow through is longer than
the instability growth time.

The growth rate and fastest-growing wavevectors are roughly
independent of the magnetic field strength, provided the magnetic
pressure is at least comparable to the gas pressure so that the fields
resist bending by gas pressure forces \citep{bs03}.  The fields serve
only to direct the fluid displacements away from the wavefront normal,
giving the gas velocity a component parallel to the radiative flux
perturbations so the radiation forces accelerate the motions making up
the mode.  Disturbances reaching large amplitude become trains of
shocks with maximum strength limited by the stiffness of the field.
With the relatively small domain used in our calculations, the effects
of the boundaries become important once the field lines bend.  The
low-density gaps opened by the photon bubbles could serve as seeds for
a large-scale Rayleigh-Taylor type instability of the kind modeled by
\cite{kk05}.  However if the magnetic pressure is less than the gas
pressure, the field bends before the density excursions become large
\citep{tb05}.  Photon bubbles will saturate at low levels in envelopes
with weak fields.  The magnetic field strength affects the escape of
stellar radiation from the birth cloud and therefore possibly the
final mass of the star.

Throughout this paper we have focused on the application of photon
bubbles to the dusty envelopes surrounding massive stars.  We note,
however, that dusty optically thick gas is present in a variety of
other astrophysical environments.  In particular, in the central $\sim
100$~parsecs of luminous starbursts such as Ultraluminous Infrared
Galaxies (ULIRGs), the entire interstellar medium is optically thick
to far infrared radiation (e.g., Downes \& Solomon~1998, Thompson et
al. 2005).
Although the magnetic field strength in this environment is uncertain,
it is likely that the magnetic energy density is comparable to or
larger than the radiation energy density, both of which are
significantly larger than the gas energy density (see Thompson et
al. 2006
for arguments to this effect based on the FIR-radio correlation).
This implies that most of the interstellar medium in ULIRGs is
unstable to photon bubbles.  The same conclusion probably applies to
the even denser gas on parsec scales around active galactic nuclei
(AGN).  Nearly half of Seyfert 2s have obscuring columns
$>10^{24}$~cm$^{-2}$ (e.g., Risaliti et al. 1999),
which corresponds to a FIR optical depth greater than unity.  Also,
models of the growth of massive black holes at high redshift imply
that much of that growth occurs with the central black hole surrounded
by an optically thick dusty envelope from $\sim 1-100$~pc (e.g.,
Hopkins et al. 2006).
Just as in the massive star problem, photon bubbles may allow the AGN
radiation to escape more readily, minimizing the impact of the
radiation pressure on surrounding gas.  In addition, if photon bubbles
lead to conditions conducive to masing around massive stars, as we
suggest below, they may also be relevant for the much more luminous
masers present in ULIRGs and AGN (mega-masers; see Lo 2005
for a review).

\section{CONCLUSIONS\label{sec:conc}}

The optically-thick dusty envelopes around young massive stars are
subject to the photon bubble instability according to linear analytic
and non-linear numerical radiation MHD calculations.  Instability
occurs if the radiation and magnetic pressures are greater than the
gas pressure and the envelope is optically-thick to its own thermal
photons.  The instability has not been seen previously in this context
because few calculations including both magnetic fields and radiation
were made in two or three spatial dimensions.  Photon bubbles develop
from 1\% density perturbations into trains of shocks within
1000~years.  Growth is fastest where the magnetic field lines are
parallel to the radiation flux vector, as can occur at the magnetic
poles or near the edges of an outflow cavity.  The shock velocity
jumps grow with time, reaching several km~s$^{-1}$ in our calculations
before the boundary effects become important and the runs are ended.
The velocity jump at which the magnetic fields bend is approximately
the Alfven speed and is similar to the mean line width of
6~km~s$^{-1}$ in hot molecular cores \citep{se03}, given field
strengths in the range 0.01--0.6~Gauss measured in water maser
emitting regions \citep{fg89,st01,st02,vd06}.  During the growth, the
field strength is largely unaffected by the shocks because the gas is
compressed along the field lines.  The shocked gas can be more than
five times denser than upstream and the temperature of 1000~K is in
the range required for maser emission in non-dissociative shock models
\citep{kn96}.  The fronts are stable in three dimensions \citep{tb05}
and will remain planar over length scales smaller than the curvature
of the magnetic field lines, so that unless the fields are tangled
with radii of curvature smaller than 10~AU, the shock trains provide
conditions suitable for water maser emission.  The lines of sight with
greatest optical depth are parallel to the shock fronts and inclined
with respect to the magnetic fields, in contrast with generic
mildly-supersonic, strongly-magnetized turbulence where the maser
emission is beamed along the fields \citep{ww04}.

Although the mean flux exceeds the Eddington limit after shocks
develop, almost all the gas remains in the domain until the magnetic
fields buckle.  The domain-integrated radiation force stays low
because the flux is greatest where the density and hence the radiation
cross-section per unit volume is least \citep{be01}.  The low-density
gas is not blown out by the radiation but remains with the
high-density gas because the two are tied together by the stiff
magnetic fields.  By making accreting material porous to radiation,
the photon bubble instability will allow stars to grow above the mass
limit imposed in spherical symmetry by the radiation force on the
dust.

The vigorous growth of photon bubbles leads to the rapid destruction
of the simple initial conditions used in our calculations.  The shock
strength increases with the separation of the fronts, which is limited
by the small size of the domain.  On larger scales the shock pattern
is likely to be disrupted once the spacing reaches the radius of
curvature of the magnetic field.  Further investigation of the effects
of the instability above the Eddington limit will require calculations
covering a larger portion of the circumstellar envelope.

\acknowledgements
  We thank the staff of the Kavli Institute for Theoretical Physics
  for their hospitality.  The KITP is funded by the National Science
  Foundation under grant PHY99-07949.  Part of the work was carried
  out at the Jet Propulsion Laboratory with support from the internal
  Research \& Technology Development program.  JPL is operated by the
  California Institute of Technology under contract to NASA.  E.~Q.\
  was supported in part by NASA grant NNG05GO22H.


\end{document}